\documentclass[a4paper,10pt]{article}
\usepackage{amsmath}
\usepackage{amsfonts}
\usepackage{amssymb}
\usepackage{graphicx}
\begin{document}

\title{  \textsc{Curvaton  and quantum gravity effect on the tensor to scalar ratio  of the chaotic inflation}}

\author{P  K  Suresh $\footnote{e-mail: sureshpk@uohyd.ac.in}$ \\[.2cm]
{\small  \it School of Physics, University of Hyderabad. }\\
{\small  \it P. O. Central University, Hyderabad 500046. India.}}

\date{\empty}

\maketitle

\begin{abstract}
The expected tensor-to-scalar ratio estimate of the upcoming CMB mission probe measurements may establish a lower value of the ratio than the currently obtained value.  It can be described in terms of a single field chaotic inflation model along with the curvaton or quantum gravity or their combined effect.   Consequently, the role of quantum gravity or curvaton in the dynamics of the early universe may not be ruled out. The curvaton scenario and quantum gravity effect can be tested experimentally.
\end{abstract}

 \section{Introduction}
The single scalar field-driven inflationary scenario has the potential to address some of the shortcomings of the standard model of cosmology \cite{guth1,lindebook}. But,  the recent estimate of the tensor-to-scalar ratio from various CMB anisotropy measurements disfavour the most popular single scalar field-driven chaotic inflationary models. However, there have been attempts to rescue the models by including quantum gravity \cite{xc} or some pre-inflationary effect \cite{sdas}. On the other hand, in absence of an integrated theory of inflation and high energy physics, some scenario predicts the existence of multi-scalar fields in the early universe and one among them of it could be a candidate of inflation, the inflaton. In the multi-field scenario, it is expected that apart from the inflationary scalar field some other scalar field(s) also responsible for the generation of primordial perturbations. The additional scalar field that contributes to the primordial cosmic perturbations is known as curvaton \cite{dh}-\cite{dhlc}.  The curvaton field is considered as sub-dominant over the inflaton and its energy density is extremely small compared to the total energy density during the inflation. However, its energy fraction may grow in the post-inflation era and can generate curvature perturbation due to its decay process. The relevance of the curvaton field along with chaotic inflation is examined with the BICEP2 and Planck results and showed that its impact can reduce the tensor-to-scalar ratio of the chaotic inflationary models compared to their standard value \cite{en,fu}.

Recently it is shown that the quantum gravity effect on the chaotic inflation can also bring down the tensor-to-scalar ratio compared to its standard value \cite{xc}. The tensor-to-scalar ratio is a useful quantity in exploring the quantum gravity via CMB through inflation. The quantum gravity effect on the inflation field is possible provided its energy much higher than the Planck scale. With such a high energy scale, the field that responsible for inflation is sensitive to quantum gravity.  Determination of the tensor-to-scalar ratio may help understand  quantum gravity  experimentally even though there is a lack of consistent quantum theory of gravity at present. However, the effective theory approach is found useful in dealing with quantum gravity and shown that the tensor-to-scalar ratio of the chaotic inflation is lower than its standard value due to the quantum gravity effect.

 The validation of the curvaton and the quantum gravity scenario is crucially dependent on the measurement of the amplitude of the primordial tensor power spectrum. In other words, the determination of the tensor-to-scalar ratio is essential in establishing the inflationary model, curvaton model, and quantum gravity. 
The measurement of very high precision data of the CMB  B-mode polarization due to the primordial gravitational waves can put stringent constraints on the model of single scalar field inflation, curvaton,  and quantum gravity.

 In absence of the curvaton and quantum gravity effect, the tensor-to-scalar ratio of the quadratic inflation is much higher than the value obtained from the  CMB observations \cite{bkp}. Nevertheless, it can be shown that the quantum gravity and curvaton effect resulting in bring the value of the ratio down, so that the single scalar inflation model may be compatible with the recent joint  CMB data analysis \cite{2018}. 
However, the expected  tensor-to-scalar ratio estimate of the upcoming CMB mission indicates that the ratio can be probed even further down \cite{ltbrd}. Hence, if the future CMB mission measurements establish 
a lower value of the ratio than the current estimated value, then a mechanism is indeed indispensable to account for it.
We show that the outlined tensor-to-scalar ratio estimate of the upcoming CMB mission may be described in terms of the single field quadratic chaotic inflation along with either the curvaton or quantum gravity effect.
 
Therefore, given the planned tensor-to-scalar ratio probe of the future CMB   B-mode mission such as  LiteBIRD, etc., we examine whether the quantum gravity and curvaton or their combined effect played any role in the dynamics of the early universe or not. Further, we seek the possibility of testing the curvaton scenario and quantum gravity experimentally through the CMB upcoming mission.

Throughout we follow the unit $c=\hbar= G=1$.
   
\section{Chaotic inflation and curvaton}

In a single field  slow-roll  inflationary scenario, the potential for chaotic inflation can be written as
\begin{equation}
V(\phi)=\frac{\phi^p} {p \,m_{pl} ^{p-4}}
\end{equation}
where $p$ is a dimension free constant. The  first and second slow-roll parameter are respectively  given by
\begin{equation}
\epsilon =  \frac{m_{pl}^2}{2}\left(\frac{V'}{V}\right)^2, \label{e}  
\end{equation}
\begin{equation}
 \eta =  m^2_{pl}\left(\frac{V''}{V}\right) \label{et},
\end{equation}
and are satisfy the conditions  $ \epsilon, \mid  \eta \mid  \ll $ 1.
   Using the  condition to terminates inflation i.e., $ \epsilon(\phi_{end}) =1,$  the e-folding number $N$  can be expressed as
\begin{equation}\label{efold}
N \simeq \frac{\phi^2}{ 2 \, p \,m_{pl}^2}- \frac{p}{4}.
\end{equation}
Hence, the slow roll parameters can be written as 
\begin{equation}\label{ep}
\epsilon = \frac{p}{4 N +p},
\end{equation}
\begin{equation}
 \eta = \frac{2(p-1)}{4 N +p}.
\end{equation}

In the standard scenario, the inflation is due to the sole effect of a single scalar field. However, it is possible to exist several fields during the inflation and at the same time which had no role in the inflation, but some of them can be  significant  in the post-inflation era.  Such fields are in general termed as curvaton ($\sigma $). In the simplest potential for a curvaton scenario can be taken as quadratic form as follows

\begin{equation}
 V(\sigma) = \frac{1}{2} m_{ \sigma}^2 \sigma^2,
\end{equation}
where $ m_{\sigma}$ is the mass of the curvaton field.

During the inflationary period, the curvaton field is massless in the sense $ m_{\sigma} \ll H$ (where $H$ is the Hubble parameter), however, it acquires mass after the termination of the inflation when  $m_{\sigma}\simeq H$. At a later stage, the curvaton field becomes dominant over the radiation because the energy density of the curvaton field,  $\rho_{\sigma}$ change according to  $ a^{-3}$  fashion of the scale factor $ a$. Whereas the energy density of the radiation, $\rho_{\gamma}$ change as $ a^{-4}.$ Hence,  the total energy density  $\rho$ of the universe becomes the subdominant of   $\rho_{\sigma}$,  the energy density of the curvaton, sometime later in the evolution of the universe.   
 
In addition to adiabatic perturbations due to the inflation, the quantum fluctuations associated with the curvaton field in the course of the inflationary phase can also lead to primordial perturbations, $ {\zeta},$ after the horizon exit. Hence the curvaton can play a role at the end of the inflationary epoch.  At the end of the inflationary era, the quantum fluctuations of curvaton also contribute to the primordial perturbations. Thus the fluctuations due to inflation get enhanced by the curvaton effect. The resulting primordial perturbations can be characterised  by the power spectrum, given by \cite{en,kl}
 
 \begin{equation}
P_{\zeta} = P_{\zeta} ^{\phi}+P_{\zeta} ^{\sigma} = P_{\zeta} ^{\phi} (1+\kappa)
\end{equation}
 where $ \kappa = \frac{P_{\zeta} ^{\sigma}}{P_{\zeta} ^{\phi}} $ is the  ratio of the curvaton  power spectrum to the inflaton power spectrum. The scalar inflaton power spectrum   $ P_{\zeta} ^{\phi}$  can be written as
\begin{eqnarray}
P_{\zeta} ^{\phi}&= & \frac{H_{*}}{8 \pi^2  \epsilon m^2_{pl}},
\end{eqnarray}
and the curvaton  power spectrum $P_{\zeta} ^{\sigma}$ is given by
\begin{eqnarray}
P_{\zeta} ^{\sigma}&= & \frac{H_{*} \beta_\sigma}{9 \pi^2  \sigma^2 _{*} m^2_{pl}},
\end{eqnarray}
here $*$  means the  quantity evaluated just before the horizon exit and  $ \beta_\sigma = \frac{\rho_\sigma}{\rho},$  when the curvaton begins to decay.

 The inflation also generates tensor perturbations due to zero-point vacuum fluctuations. Therefore,  the tensor-to-scalar ratio with curvaton power spectrum can be written as
\begin{equation}\label{r}
r \equiv \frac{P_{\cal T} }{P_{\zeta} }  = \frac{16\epsilon}{1+\kappa}.
\end{equation}

Similarly, the scalar  spectral index can be expressed  with the curvaton as 
\begin{eqnarray}\label{nc}
n_s-1&=&\frac{d \ln P_{\zeta}}{d \ln k}=  2 \eta_\sigma - 2\epsilon - \frac{4 \epsilon - 2 \eta }{1+\kappa}, 
\end{eqnarray}
where $\eta_\sigma = \frac{m_\sigma ^2}{3 H_* ^2}.$ 

Using (\ref{ep}), (\ref{r}) and (\ref{nc}), we get 

\begin{eqnarray}\label{mr}
r= 4 p\left ( 1- n_s +  2 \eta_\sigma  -  \frac{2p}{4 N +p}\right).
\end{eqnarray}

When $p=2$ and $p=4$ the result (\ref{mr}) respectively  leads to  the quadratic and quartic inflationary model case. The present study is focusing on quadratic chaotic inflation only.

\section{Quantum gravity  and chaotic inflation}

In the effective theory  approaches to quantum gravity,  the potential  for the inflaton is given by 
\begin{equation}
\label{ }
V(\phi)= V_{ren}(\phi )+ \sum_{n=5}^{\infty} c_{n} \frac{\phi^n}{m_{pl}^{n-4}}.
\end{equation}
Where $ V_{ren}(\phi )$  contains all the renormalizable terms up to four dimensions and the second term is due to quantum gravity.  Here $c_{n}$  are the Wilson coefficients and its value must be the order of 10$^{-3},$ so that the inflaton potential remains as flat \cite{rh}.  For a chosen inflationary potential one specific dominant term of the Wilson coefficients only is considered and the remaining are ignored.  The expression $c_{n} \frac{\phi^n}{m_{pl}^{n-4}} $  is known as   higher dimension operator (HDO).

The potential  for the quadratic chaotic inflation  can be written   in the effective theory  as \cite{xc}
\begin{equation}\label{cHDO}
V(\varphi)=m_{pl}^4\left( \bar{m}^2 \varphi^2+ c_n\varphi^n \right),
\end{equation}
where $\bar{m}= m/m_{pl} $ and  $ \varphi= \phi/m_{pl}$   are the  normalized mass  and  inflation field   with  respect to the reduced Planck mass.   In the absence of  the  HDO,   (\ref{cHDO}) reduces to the standard quadratic chaotic inflationary potential. 

Since the HDO should be a correction to the leading term on $\bar{m}^2 $  \cite{xc},  we consider the relevant  term
$c_6 = \alpha_m \bar{m}^2. $  Hence  the potential (\ref{cHDO}) can be recasted   as
\begin{equation}
\label{cHDOr}
V(\varphi)= m_{pl}^4 \bar{m}^2 \varphi^2\left(1+ \alpha_m \varphi^4 \right),
\end{equation}
with the  condition that 
\begin{equation} \label{cnd}
 \mid  \alpha_m  \mid \varphi^4 < 1.
 \end{equation}

Therefore  by ignoring the higher order terms in the potential, the first and second slow parameters for  the  quadratic chaotic inflation potential   with quantum gravity effect, we get  
\begin{equation}
\label{eh1}
\epsilon=\frac{2 m_{pl}^2}{\varphi^2} \left(  1+ 4\alpha_m \varphi^4\right),  
\end{equation}
\begin{equation}
\label{et1}
\eta= \frac{2 m_{pl}^2}{\varphi^2} \left(  1+14 \alpha_m \varphi^4\right). 
\end{equation}
Using the condition to end the inflation, the slow-roll parameters can be expressed in terms of the e-folding number. Hence  the  tensor-to-scalar ratio  of quadratic inflation with quantum gravity effect in terms of the scalar spectral index is obtained as
\begin{equation}
\label{ }
r= 8\left [ 1- n_s - \frac{1}{N} \left(1-   \frac{  N^2 \alpha_m}{12  \pi^2} \right) \right ].
\end{equation}
According to the effective theory, the value of the HDO parameter of the quadratic  inflation must be   $ {\mid \alpha_m \mid}^{\tiny EFT}  \lesssim  2 \times 10^{-2}. $

\begin{figure}[]
\centering
 \includegraphics[scale=0.38]{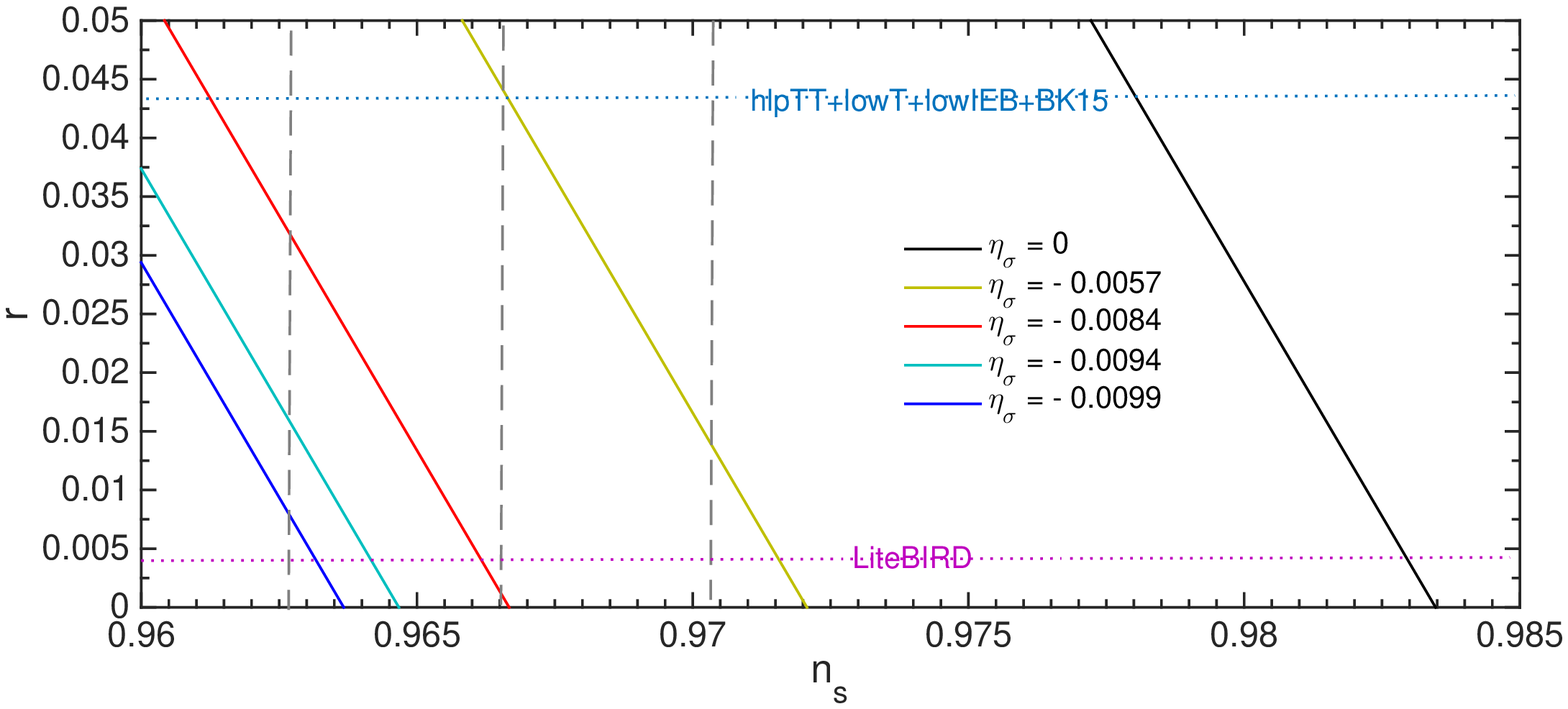}
 \centering
\caption{ Curviton effect (solid colour  lines)  on the tensor-to-scalar ratio  for  the quadratic chaotic inflation  for  various values the curvaton parameter   with  the hlpTT+lowT+lowIEB+BK15   and the  LiteBIRD   upper bound. The dashed grey lines are the  bound on $n_s$ from the Planck mission \cite{2018}. \label{f1}}
\end{figure}

\begin{figure}[]
\centering
 \includegraphics[scale=0.38]{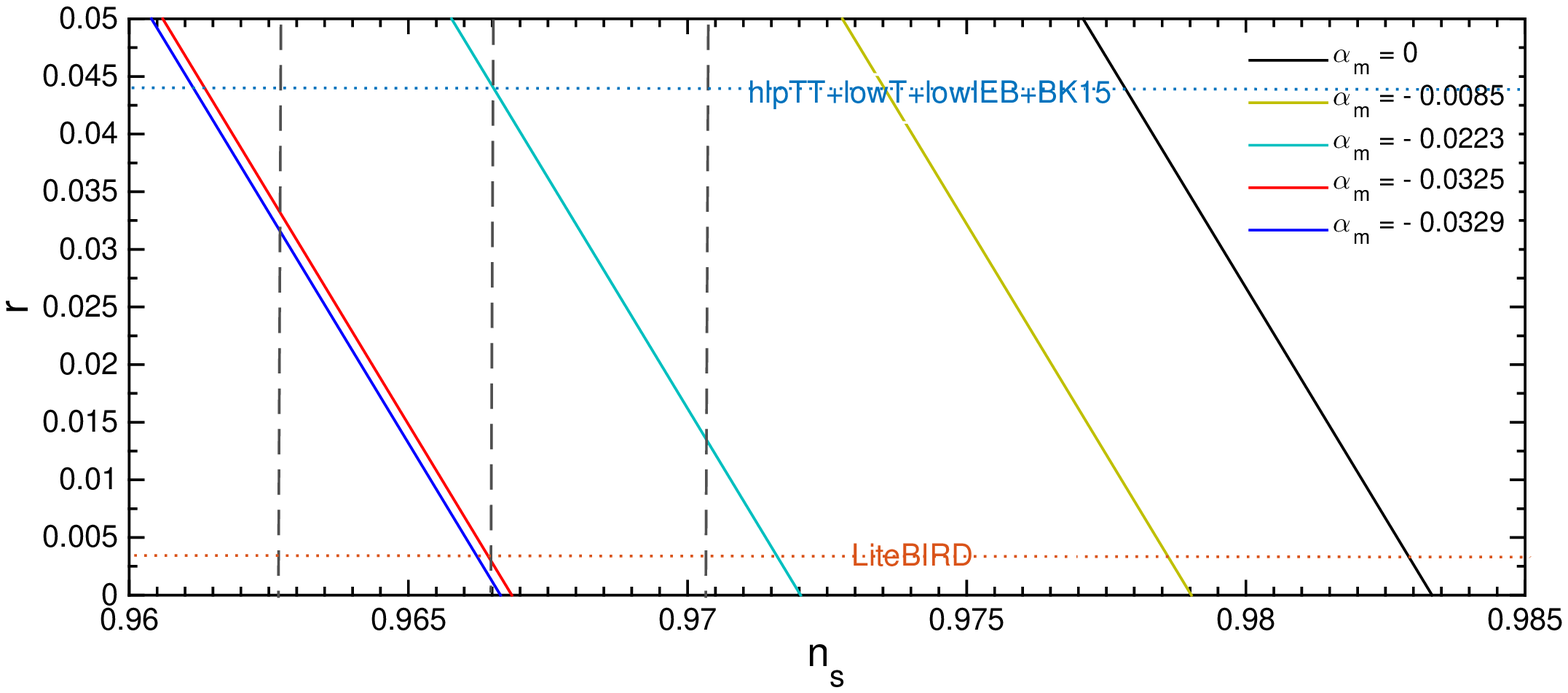}
 \centering
\caption{ Quantum gravity effect (solid colour  lines)   on the tensor-to-scalar ratio  for  the quadratic chaotic inflation  for  various values the HDO parameter   with  the hlpTT+lowT+lowIEB+BK15   and the  LiteBIRD   upper bound. The dashed grey lines are the  bound on $n_s$ from the Planck mission \cite{2018}. \label{f2}}
\end{figure} 

\begin{figure}[]
\centering
 \includegraphics[scale=0.35]{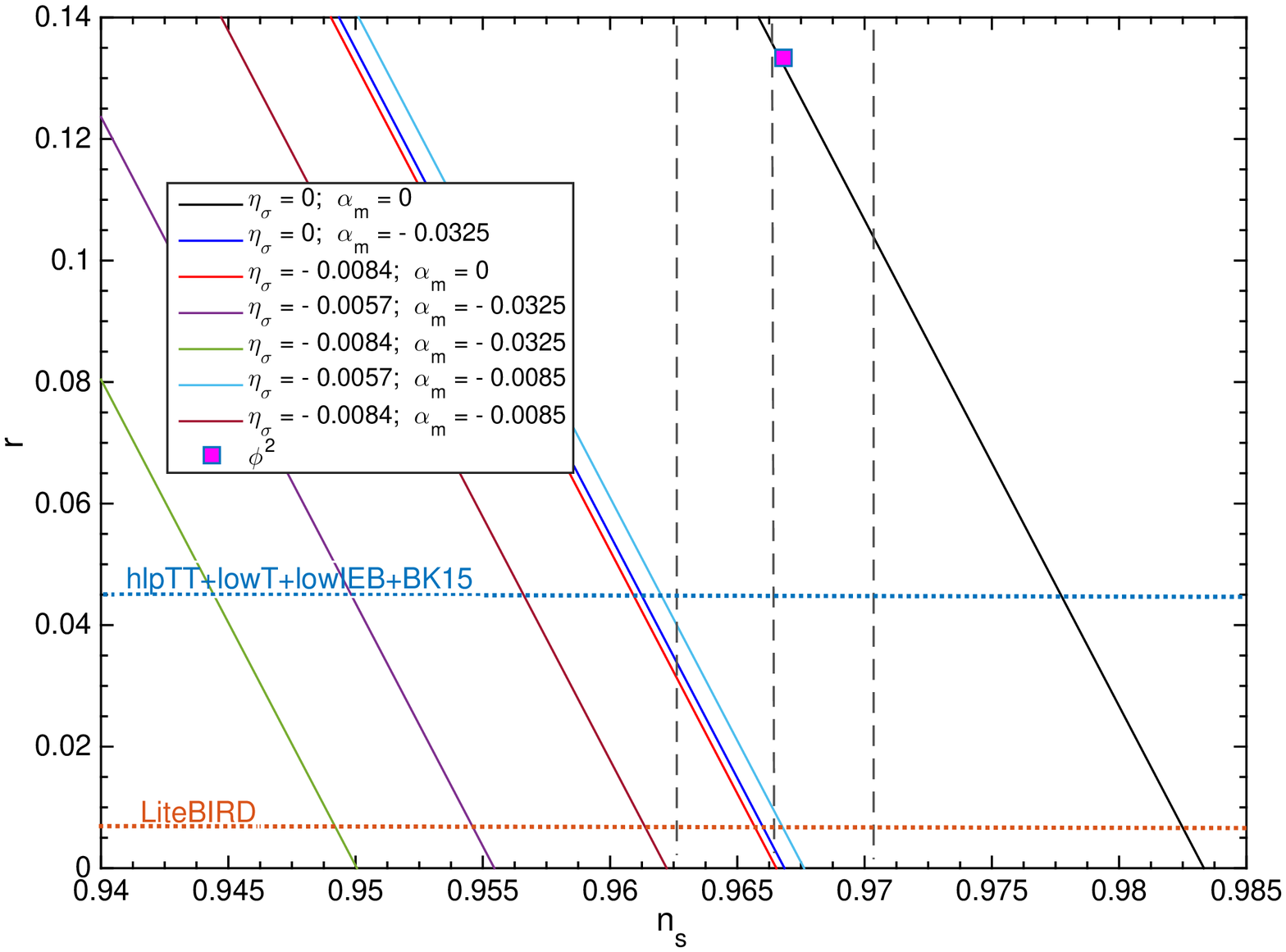}\\
 \includegraphics[scale=0.35]{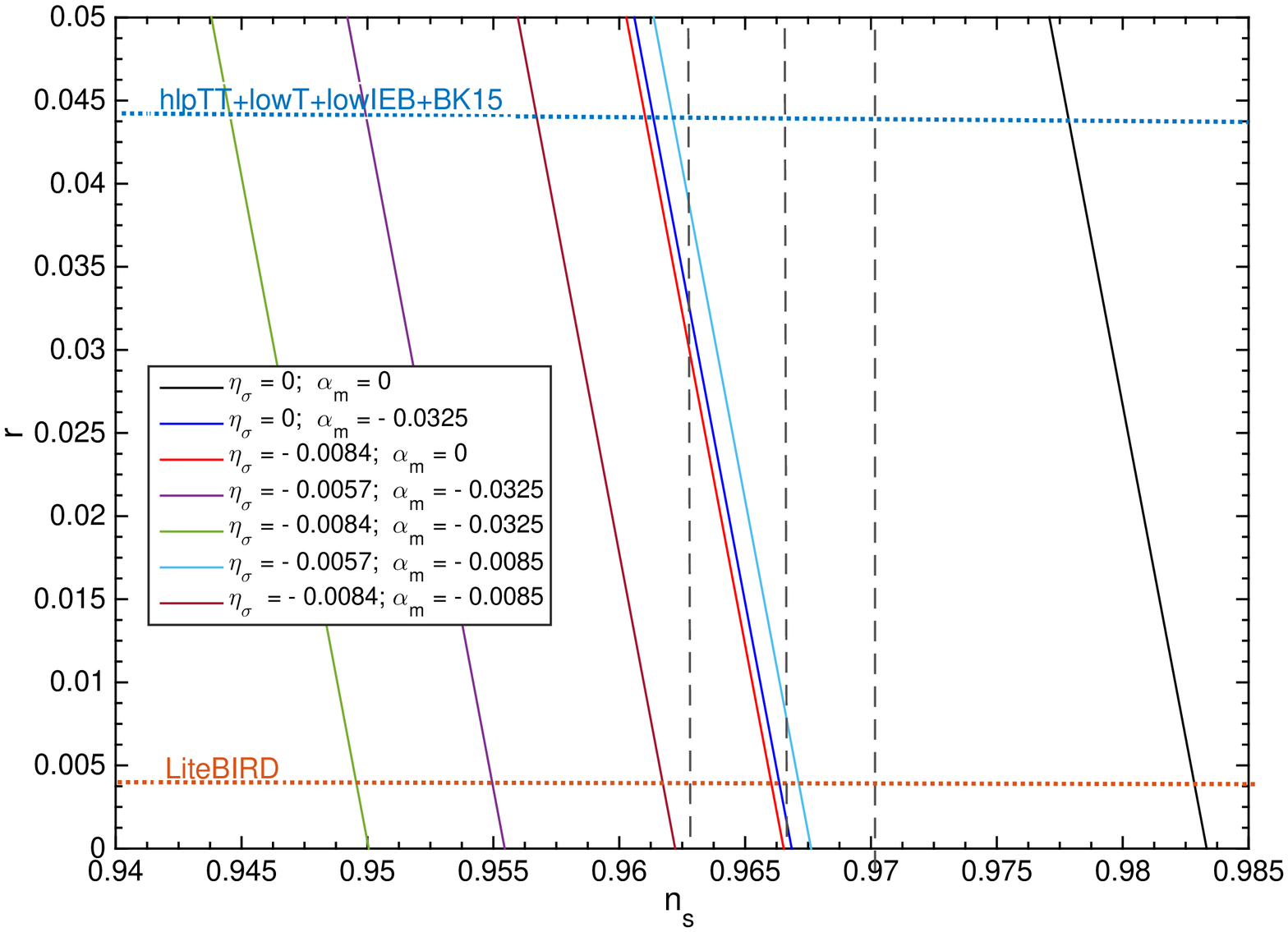}
 \centering
\caption{ The combined effect of curvaton and quantum gravity (solid colour lines)  on the tensor-to-scalar ratio for the quadratic chaotic inflation for various values the curvaton and HDO parameter with the hlpTT+lowT+lowIEB+BK15   and the  LiteBIRD upper bound. The dashed grey lines are the bound on $n_s$ from the Planck mission \cite{2018}. The upper panel includes the standard quadratic inflation case $r$  value.  The lower panel is highlighting the hlpTT+lowT+lowIEB+BK15   and the  LiteBIRD cases. \label{f3}}
\end{figure}

\section{Joint effect of  curvaton  and  quantum gravity  on the chaotic inflation}
By considering the curvaton scenario with the quantum gravity effect, the tensor-to-scalar ratio for the quadratic inflationary model is obtained in terms of the scalar spectral index as
\begin{equation}
\label{ }
r= 8\left [ 1- n_s - \frac{1}{N} \left(1-   \frac{  N^2 \alpha_m}{12  \pi^2} \right) +2 \eta_\sigma \right ].
\end{equation}

 In absence of the curvaton and quantum gravity effect, the tensor-to-scalar ratio of the quadratic inflation model takes its standard value, but it is much higher than the estimated value from the  CMB observation \cite{bkp}. Nevertheless, including additional effects such as the quantum gravity and curvaton can bring down the value of the tensor-to-scalar ratio of the quadratic inflation, so that the model may be still retained with the recent estimate of the hlpTT+lowT+lowIEB+BK15 joint data \cite{2018}.
However, the upcoming CMB mission is planned to probe the value of the ratio even further down \cite{ltbrd}. Therefore, by taking into account the expected estimate of the tensor-to-scalar ratio of the upcoming CMB mission,  the LiteBIRD, we estimate the ratio for the quadratic chaotic  inflation for various values of the curvaton parameter and HDO parameter as well as their joint effect. The value of the HDO parameter is obtained within the range of the tensor-to-scalar ratio,  $ r \in [10^{-3}, 0.1],$  but subjected to the EFT upper bound. Similarly, the value of the $\eta_\sigma$ also calculated with the aforementioned range of the tensor-to-scalar ratio.

We estimate the tensor-to-scalar ratio for various values of the curvaton parameter for  the spectral index $n_s\in$ [.96,.985] and the obtained results are presented in Fig (\ref{f1}). The estimated ratio for various values of $\eta_\sigma$ is compared with the result of the hlpTT+lowT+lowIEB+BK15 joint data as well as the expected probe of the  LiteBIRD.  The result shows that the curvaton effect is responsible  for  lowering the tensor-to-scalar ratio of the quadratic chaotic inflation. Further, either the current estimate or the upcoming probe range is not ruling out the curvaton scenario. 

We  estimate the tensor-to-scalar ratio for various values of the HDO parameter for  the scalar spectral index $n_s\in $ [.96,.985] and  results are given in Fig (\ref{f2}). The estimated ratio for various values of $\alpha_m$ is compared with the result of the hlpTT+lowT+lowIEB+BK15 joint data and the expected probe of the  LiteBIRD. The result shows that the quantum gravity effect brings down the tensor-to-scalar ratio of the quadratic chaotic inflation. It can be observed that neither the current estimate nor the LiteBIRD probe range disfavour quantum gravity.

Further, we also study the joint effect of curvaton and quantum gravity on the tensor to scale ratio for the scalar spectral index range $n_s\in$ [.94,.985]. The estimated results for various values of the curvaton and HDO parameter are compared with the hlpTT+lowT+lowIEB+BK15 joint data and the future probe of the LiteBIRD. The obtained results are depicted in Fig (\ref{f3}). The analysis of joint effect results of the curvaton and quantum gravity indicate that neither the current estimate nor the upcoming LiteBIRD probe range disfavour the aforementioned scenario.

\section{Conclusion}
In the standard scenario,  the tensor-to-scalar ratio of the quadratic chaotic inflation is much higher than the estimate of various CMB measurements, hence currently the model is considered unfavourable. Nevertheless, the inclusion of additional phenomena such as the quantum gravity or curvaton effect may bring down the value of the ratio, thereby the model may be shown compatible with the recent  CMB joint data analysis. 
However, the expected tensor-to-scalar ratio estimate of the upcoming CMB mission probe may measure it further down. Hence, if the future CMB mission measurements establish a much lower value of the ratio than the current estimated value, then a mechanism is required to account for it.
The present study shows that the outlined tensor-to-scalar ratio estimate of the upcoming CMB mission result may be described in terms of the single field quadratic chaotic inflation along with the curvaton or quantum gravity or their combined effect. The outcome of the study is mainly based on the analysis of the tensor-to-scalar and scalar spectral index plane and showed that the quantum gravity and curvaton or their combined effect probably played a vital role in the dynamics of the early universe. 

In the $ (r,n_s)$ plane the curvaton and the quantum gravity effects may be degenerated and hence may be difficult to distinguish between them. However, the future observations and determination of  $n_s$   and  $r$  value may resolve the situation.  Hope that the upcoming high-quality precision data of CMB  may validate the  scenarios unambiguously.

\end{document}